\documentclass[letterpaper, 10 pt, journal, twoside]{IEEEtran}
\usepackage[pdftex]{graphicx} 
\usepackage{gensymb}
\usepackage{amsmath}
\usepackage{hyperref}
\IEEEoverridecommandlockouts  
\usepackage{booktabs}
\usepackage{xcolor}
\usepackage[normalem]{ulem}
\usepackage{tikz}
\usepackage{xfrac}
\usepackage{float}
\usepackage{placeins}
\usepackage[justification=centering]{caption}
 \usepackage[pscoord]{eso-pic}
 \newcommand{\placetextbox}[3]{
 \setbox0=\hbox{#3}
 \AddToShipoutPictureFG*{ \put(\LenToUnit{#1\paperwidth},\LenToUnit{#2\paperheight}){\vtop{{\null}\makebox[0pt][c]{#3}}}
 }
 }
 \placetextbox{.23}{0.055}{\small979-8-3503-0965-2/23/\$31.00~\copyright 2023 IEEE}
\begin{document}

\title{Predicting Dosage of Immunosuppressant Drugs After Kidney Transplantation Using Machine Learning
\thanks{* Authors Contributed Equally}}
\author{
\begin{tabular}{c} Kapil Panda* \\ University of North Texas \\ Denton, United States of America \\ kapil.panda30sc@gmail.com \end{tabular} \and
\hspace{50pt} % Change this number to change the space between authors
\begin{tabular}{c} Anirudh Mazumder* \\ University of North Texas \\ Denton, United States of America \\ anirudhmazumder26@gmail.com \end{tabular}
}
\maketitle
\begin{abstract}
While kidney transplants are seen as the best treatment option for patients with end-stage renal disease and kidney failure, the organ's health depends on the dosage of immunosuppressant drugs post-transplantation. Due to the dosage variance based on each patient's unique physiology, nephrologists face numerous difficulties when determining the precise dosage needed for each patient. Therefore, in this research we aim to devise a machine learning algorithm to forecast the dosage of immunosuppressant drugs needed for different patients after kidney transplantation. Utilizing a random forest algorithm, the devised model is able to achieve accurate measurements for patient drug dosages.
\end{abstract}


\section{Introduction}
Kidney transplants are widely viewed as the best treatment for end-stage renal disease and kidney failure, offering patients the chance to resume a healthy, normal lifestyle. The success and longevity of these transplants, however, depends on the effective management of immunosuppressant drugs post-transplantation \cite{Muduma2016}. After receiving a new kidney, the recipient's immune system recognizes it as a foreign organ and may trigger an immune response to reject the kidney\cite{Halloran2004}. Immunosuppressant drugs, on the other hand, suppress the recipient's immune system to prevent this rejection process and minimize the risk of infections and other complications. Therefore, ensuring proper and consistent immunosuppressant drug administration is essential to prolong its functionality and demands meticulous dosage specifications to balance the medication's efficacy and adverse effects. Achieving this optimal dosage, however, is challenging due to the various patient-specific factors that play a role, making it an intricate puzzle for nephrologists to solve. Given the lack of kidneys available for donation, it is all the more necessary to handle every transplantation with the utmost care \cite{Kamal2022}.

In recent years, the progress made in artificial intelligence and machine learning has opened up new possibilities in healthcare for both detection and prediction. With the incredible capabilities of AI and ML algorithms and models, healthcare professionals now have powerful tools that can aid in analyzing vast amounts of medical data with unparalleled precision and efficiency, thereby increasing the efficacy of medical prescriptions \cite{Reddy2020}. However, certain domains have yet to be completely immersed in the frontiers of AI and ML, with kidney transplantations, specifically medication dosage, being one of them.

Therefore, in this research we aim to leverage the power of machine learning and develop a novel algorithm to help predict the exact dosage of immunosuppressant drugs to be administered after kidney transplantation. To do this, a Random Forest Regression algorithm is utilized, that is trained on a diverse subset of data containing various features and data points, allowing the combination of predictions from multiple decision trees. Consequently, the algorithm achieved robust and accurate predictions, indicating the viability of the model.
\section{Methodology}
\subsection{Materials}
A dataset containing data about immunosuppressant prescriptions in kidney transplantation was utilized to train the algorithm and can be seen at \cite{Chang2017}. Additionally, Python was used in order to write and compile the code.
\subsection{Algorithm}
\subsubsection{Data Preprocessing}
In order to utilize the data effectively for training the machine learning model, it was crucial to perform data parsing to extract relevant features \cite{Famili1997}. The dataset was divided into critical variables, such as age, gender, height, weight, and other parameters related to the patient's pre-transplant situation. However, there were instances of missing data, which needed to be addressed to ensure the integrity of the dataset. To handle this, missing values were imputed by replacing them with the average value of the corresponding column. This step was essential to avoid any biased outcomes due to incomplete data \cite{Patrician2002}.

Additionally, specific categorical data, which could not be directly converted into a numerical form using one-hot encoding, had to be dropped from the dataset \cite{Okada2019}.  
This decision was imperative as the machine learning algorithm used for the task was a regression predictor, which requires numerical input and output for accurate processing and analysis \cite{Nunez11}. By retaining only the relevant numerical data, the machine learning model could effectively learn from the dataset and generate meaningful predictions.

The careful preparation and preprocessing of the data were fundamental to the success of the machine learning model. These steps allowed for building a robust and accurate predictor for kidney transplant outcomes, enabling a better understanding of patient outcomes and contributing to improving kidney healthcare. The model's ability to process numerical data and handle missing values efficiently ensured that it could make reliable predictions based on the available information, offering valuable insights to aid in medical decision-making and patient care.
\subsubsection{Data Analysis}
\begin{figure}[!htb]
	\centering
	\includegraphics[width=\columnwidth]{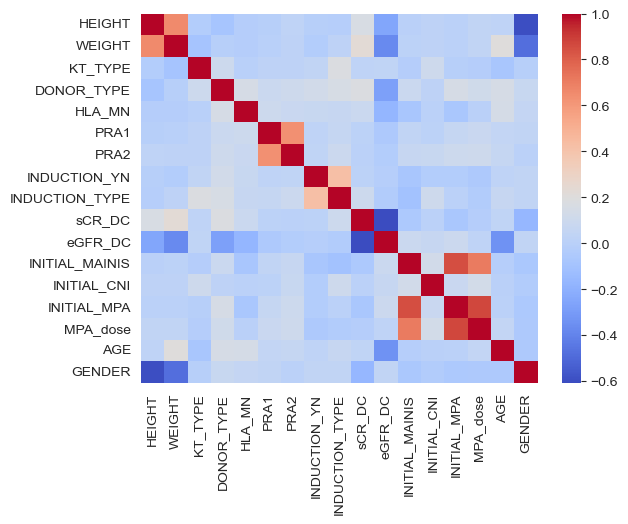}
	\caption{Correlations between each variable}
	\label{fig: Figure 1}
\end{figure}

\begin{figure}[!htb]
	\centering
	\includegraphics[width=\columnwidth]{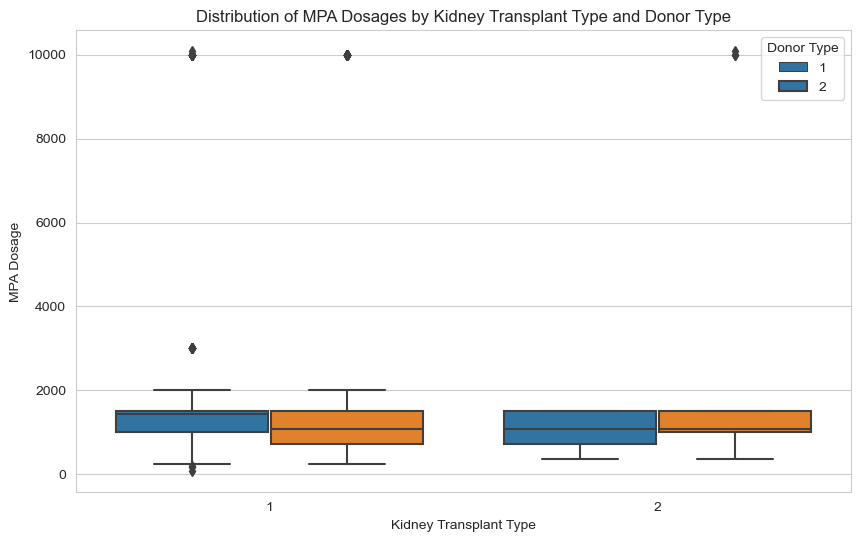}
	\caption{MPA dosages based on kidney and donor types}
	\label{fig: Figure 2}
\end{figure}

An exploratory data analysis was conducted to properly understand if there was any correlation between the dosage and the independent variables considered from the dataset. Figure \ref{fig: Figure 1} shows the correlation values between all of the variables, indicating that the initial Mycophenolic Acid (MPA) dosage and the patient's initial immunological status (INITIAL\_MAINIS) had a high correlation with the MPA dosage. The correlations show some of the features which will be highly important when creating the prediction algorithm, as they will lead to the highest accuracy by showing a high correlation.
\FloatBarrier
\FloatBarrier
Additionally, to thoroughly analyze the data, the Kidney Type and Donor Type variables were compared against the MPA dosage variable, as the immunosuppressant dosage directly impacts the kidney transplantation process and long-term viability. This comparison can be visualized in Figure \ref{fig: Figure 2}, which plots the distribution of MPA dosage values categorized by Kidney and Donor types. A key trend evident in this plot is that patients who received a kidney transplant from a living donor (Donor Type = 1) required notably higher MPA dosages overall compared to patients with deceased donors (Donor Type = 2). This correlation highlights that whether the organ donor is alive or deceased at the time of transplantation could significantly influence the optimal MPA dosage for preventing organ rejection. Specifically, recipients of kidneys from living donors exhibited a shift toward higher MPA dosage values, with a mean dosage of nearly 2000 mg for living donors versus approximately 1500 mg for deceased donors. The significantly higher dosage requirements for living donor transplants suggest these patients may undergo more aggressive immunosuppression to mitigate their elevated immunologic risks relative to deceased donor transplantation. Overall, Figure \ref{fig: Figure 2} illustrates the importance of accounting for both the Kidney Type and Donor Type factors when determining appropriate MPA dosage protocols to balance efficacy in preventing organ rejection with the minimization of adverse side effects. Further research into tailored dosage regimens accounting for these factors could enable improved clinical outcomes.
\FloatBarrier
\FloatBarrier
\subsubsection{Machine Learning}
The machine learning algorithm chosen for the research project was a Random Forest Regression algorithm, a powerful ensemble technique known for its robustness and ability to handle complex datasets effectively. Random Forest creates multiple decision trees, each trained on a random subset of the dataset, using different features and data points. This "bootstrap aggregating" process ensures that the trees are diverse, preventing them from providing the same output and reducing the risk of bias \cite{Cootes2012}. Training each tree on a subset of data and features makes the trees de-correlated, improving generalizability and reducing overfitting \cite{Matsuki2016}.

The essential advantage of using the Random Forest algorithm in this study was its capability to mitigate the impact of outliers and noise in the data \cite{Gislason2006} \cite{Agjee2018}. By aggregating the predictions from multiple decision trees and taking their average, the algorithm diminishes the influence of extreme or erroneous data points, thus resulting in more stable and reliable predictions. In contrast to relying on a single decision tree, which could be heavily affected by outliers, Random Forest's ensemble approach yields a more balanced and robust outcome. The diversity of the trees makes the overall model resilient to outliers that may skew an individual tree's predictions.

Random Forest avoids overfitting and improves generalization performance through its technique of training each tree on a subset of features. This de-correlates the trees, allowing the overall model to better generalize to new data rather than overly fitting to the training data. Using feature subsets prevents any single feature from dominating, reducing variance and enhancing predictive stability. By training trees on different feature subsets, Random Forest prevents overfitting that hurts generalizability. The resulting model is versatile and robust, able to handle new data well rather than being overly specialized to the training examples. Random Forest's ensemble approach provides inherent generalizability advantages over single tree methods.

Random Forest is an efficient and versatile algorithm owing to its ensemble approach of training multiple decision trees on subsets of data. This allows Random Forest models to leverage parallel computing for scalability while the ensemble averaging helps reduce overfitting compared to single tree methods. A key advantage is the ability to handle mixed categorical and continuous data without preprocessing, providing flexibility across diverse datasets \cite{Cutler2012}. Additionally, Random Forest performs implicit feature selection by identifying the most important variables during training, reducing dimensionality and irrelevant features automatically \cite{Chen2020}. These benefits make Random Forest well-suited for large, high-dimensional datasets for both regression and classification modeling tasks, as evidenced by its widespread adoption. The combination of computational efficiency, built-in feature selection, and applicability to diverse data types makes Random Forest an excellent general-purpose modeling algorithm.
\section{Results}
\subsection{Model Performance}
A few key metrics were calculated to evaluate the accuracy of the machine learning algorithm in predicting the dosage given to a patient after a kidney transplant. The first metric computed was the R-squared ($R^2$) value \cite{Cameron1996}. The $R^2$ metric shows the proportion of variance in the dependent variable that the independent variables can explain. The $R^2$ value ranges between 0 and 1, with a value of 1 indicating that the independent variables perfectly predict the dependent variable. A value of 0 indicates no correlation between the independent and dependent variable, while a value of 1 indicates that there is a very strong correlation between the independent variable and the dependent variable \cite{Chicco2021}.

After running the machine learning algorithm on the data, an $R^2$ value of 0.96 was achieved. This high value indicates that the independent variables utilized in the model are highly explanatory of the dosage that should be given to patients after a kidney transplant. It also shows that the model has a solid fit for the underlying data, allowing it to achieve this high $R^2$ value. The high $R^2$ implies that the dosage is strongly determined by the set of features fed into the model. The $R^2$ provides evidence that the model effectively captures the relationship between the input variables and the dosage.

In addition to $R^2$, the Mean Absolute Error (MAE) was calculated \cite{Willmott2005}. The MAE measures the absolute difference between the predicted and actual values of the target variable. A high MAE indicates the predictions are further away from the true values, implying worse predictive performance. For this machine learning model, the MAE was 245.49. The MAE is a valuable metric for evaluating the precision of the model's predictions on individual data points.

The MAE suggests that although the algorithm fits the training data well overall when making predictions on individual data points, it produces average values that are a moderate distance from the true dosage. This discrepancy could be explained by outliers, which lead to a large gap between predicted and actual dosages. The MAE reveals shortcomings in making precise dosage predictions from the features.

\begin{figure}[!htb]
	\centering
	\includegraphics[width=\columnwidth]{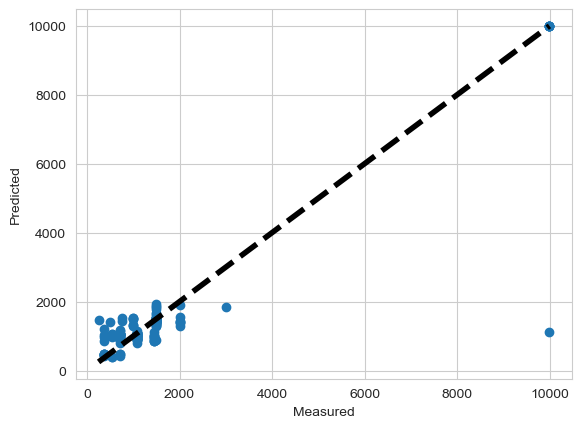}
	\caption{Predicted Output vs. Actual Output}
	\label{fig: Figure 3}
\end{figure}

Overall, while the MAE indicates the algorithm's predicted dosages deviate somewhat from the true values, the high $R^2$ shows that the model powerfully captures the relationship between inputs and output. This implies that the model is predicting well in the general sense but needs further tuning to achieve higher precision. Figure \ref{fig: Figure 3} shows the differences between actual and predicted dosage values. The $R^2$ and MAE provide complementary insights into the model's performance.

\begin{figure}[!htb]
	\centering
	\includegraphics[width=\columnwidth]{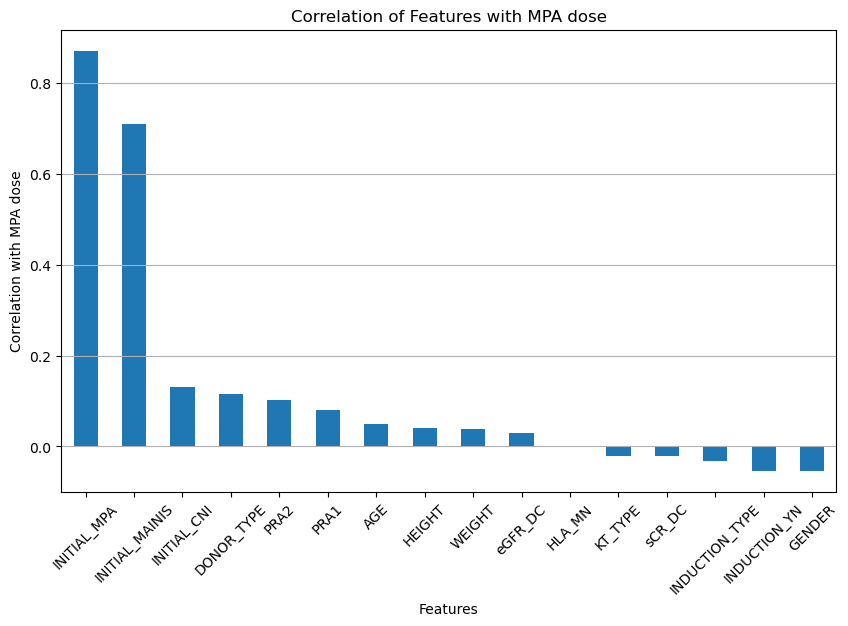}
	\caption{Feature correlation with the MPA dosage}
	\label{fig: Figure 4}
\end{figure}

To further improve the model, additional steps could be taken, such as identifying and removing outliers in the training data, trying different regression algorithms, and incorporating new features that better explain the variance in dosage. Feature engineering and hyperparameter tuning could increase precision and reduce errors. The high $R^2$ is promising, but refinements in the model could enhance its ability to make accurate predictions at the individual data point level \cite{Nishat2018}.
\FloatBarrier
\FloatBarrier
A correlation analysis was conducted to gain further insight into the relationships between input variables and the final immunosuppressant dosage \cite{Senthilnathan2019}. The Pearson correlation coefficient was calculated between each feature and the target dosage variable. As seen in Figure \ref{fig: Figure 4}, the initial MPA level exhibited the strongest positive correlation (0.87), followed by the initial main immunosuppressant concentration (0.71) and initial CNI level (0.13). This suggests that the starting MPA and primary immunosuppressant levels, measured right before dose optimization, were highly predictive of the eventual stabilized dose. Donor type, panel reactive antibody levels, age, height, and weight showed weak positive correlations ranging from 0.05 to 0.12. These donor and recipient characteristics may influence dosing to a small extent. However, the major histocompatibility complex mismatch score was not significantly correlated. Kidney type and induction therapy use also had negligible correlations, indicating they did not directly impact the dose.

Interestingly, the estimated glomerular filtration rate, serum creatinine, and gender had mildly negative correlations with the final dose. This may reflect underlying relationships between these factors and the pharmacokinetics of the immunosuppressants. Overall, the initial drug levels measured on day five post-transplant appeared to be the strongest determinants of eventual dosing, while recipient demographics and clinical variables showed minimal associations. The correlation analysis provided valuable insights into which variables were most predictive of the final optimized dosage, and these findings will help guide feature selection for the machine learning model. The variables exhibiting significant correlations with the target will be prioritized during model development, while those showing little association may not improve predictive power.

These results suggest that the starting MPA and CNI levels, along with select recipient factors like age and weight, will be essential to include as inputs to the model. However, clinical measures of kidney function and variables related to the transplant do not need to be utilized. By refining the feature set to the optimal combination of informative variables, the model can be tuned for accurate, personalized dose prediction. The correlation analysis indicated which input factors are most crucial to incorporate into the machine learning algorithm for robust dosage forecasting.
\FloatBarrier
\FloatBarrier
\section{Discussion} 
\subsection{Conclusion}
Through this research, an effective and accurate prediction model was devised to forecast the dosage of immunosuppressant drugs needed for patients after a kidney transplant with the help of machine learning. By utilizing a Random Forest Regression algorithm, the model was trained on a random subset of data with different features and data points to aggregate the predictions from all decision trees and, thus, produce a final output. As a result, the algorithm can achieve a more robust and accurate prediction while mitigating the influence of individual data points or unusual patterns in the dataset.
\subsection{Potential Implications}
This model can hold significant implications for patient care and treatment outcomes regarding kidney transplantations in patients. By leveraging this framework, clinicians and nephrologists can utilize a tool that analyzes vast amounts of patient-specific data, including genetic profiles, clinical parameters, and treatment histories, to forecast the ideal drug dosage for each recipient accurately. With precise dosage measurements of immunosuppressant drugs playing a crucial role in preventing organ rejection by suppressing the recipient's immune system, achieving an optimal dosage balance tailored to each patient's unique physiology can enhance the drug's efficacy while minimizing adverse effects. Furthermore, using this machine learning model can lead to improved patient recovery rates, reduced instances of rejection, and enhanced long-term organ viability, ultimately resulting in better patient outcomes and quality of life following kidney transplantation.
\subsection{Future Work}
In the future, we aim to utilize more comprehensive patient data to improve the accuracy and generalizability of the dosage prediction model. In this initial study, a limited set of patient variables such as age, gender, height, weight, kidney type, and donor type were utilized \cite{Chen2019}. Additionally, in follow-up work, the plan is to incorporate additional patient-specific variables, such as genetic markers, comorbidities, medications, and other clinical parameters. Expanding the feature set to include relevant genetic, physiological, and pharmacological factors could better capture inter-individual variability in immunosuppressant pharmacokinetics. Furthermore, genetic markers related to drug metabolism could provide insight into optimal dosing based on an individual's unique genotype \cite{Desai2011}.

There is also an intention to explore a longitudinal data analysis approach. Instead of considering patient data at a single time point, data could be collected over multiple time points post-transplantation. This would allow the model to capture the dynamic changes in the patient's health status, kidney function, and response to the drugs over time. Factors like weight fluctuations, evolving kidney function, and changes in concomitant drugs can impact optimal dosing. A longitudinal analysis accounting for these temporal effects could significantly improve prediction accuracy \cite{Nashif2018}.

Testing the model on more diverse, multi-center datasets will also be necessary for further validation\cite{aki}. Evaluating the performance of external populations from different clinics would assess the model's generalizability across healthcare systems and demographic groups. Access to larger-scale data reflecting varied patient subgroups could strengthen the robustness and clinical applicability of the tool.

Regarding the model itself, there is a plan to investigate using more advanced neural network architectures instead of machine learning algorithms. Complex deep-learning approaches may better handle high-dimensional clinical data and discern intricate patterns that improve accuracy. There will also be continued optimization of ensembling techniques and hyperparameter tuning to maximize predictive performance. Overall, expanded patient data, longitudinal tracking, external validation, and refinements to the model architecture and parameters hold promise for enhancing this precision dosing system \cite{Boateng2020}.
\section{Acknowledgment}
We would like to thank the University of North Texas for providing us with the resources and support to conduct this research. The invaluable guidance and encouragement from our professors and mentors have been instrumental in shaping the direction and scope of this study. We would also like to acknowledge the National Kidney Foundation for their support and inspiration in conducting this project. We would also like to thank our families for supporting us throughout our research.
\bibliographystyle{IEEEtran}

\bibliography{Bibliography}

\end{document}